\documentclass[iop]{emulateapj} 
\usepackage{natbib,amsmath,booktabs,apjfonts,threeparttable,multirow}
\usepackage{natbib}
\usepackage{graphicx}
\usepackage{rotating}
\usepackage{tabularx}
\usepackage{longtable}
\usepackage{xcolor}

\begin{document}

\title{X-RAY ANALYSIS OF THE PROPER MOTION AND PULSAR WIND NEBULA
FOR PSR J1741-2054}

\author{Katie Auchettl\altaffilmark{1,2}, Patrick Slane\altaffilmark{1},
Roger W. Romani\altaffilmark{3}, Bettina Posselt\altaffilmark{4},
George G. Pavlov\altaffilmark{4}, Oleg Kargaltsev\altaffilmark{5},
C-Y. Ng\altaffilmark{6}, Tea Temim\altaffilmark{7,8}, Martin. C.
Weisskopf\altaffilmark{9}, Andrei Bykov\altaffilmark{10}, Douglas
A. Swartz\altaffilmark{9}}

\altaffiltext{1}{Harvard-Smithsonian Center for Astrophysics, 60 Garden Street, Cambridge, MA 02138, USA}
\altaffiltext{2}{School of Physics \& Astronomy, Monash University, Melbourne, Victoria 3800, Australia}
\altaffiltext{3}{Department of Physics, Stanford University, Stanford, CA 94305, USA}
\altaffiltext{4}{Department of Astronomy \& Astrophysics, Pennsylvania State University, 525 Davey Lab, University Park, PA, 16802, USA}
\altaffiltext{5}{Department of Physics, The George Washington University,
725 21st St, NW, Washington, DC 20052}
\altaffiltext{6}{Department of Physics, The University of Hong Kong, Pokfulam Road, Hong Kong, China}
\altaffiltext{7}{Observational Cosmology Lab, Code 665, NASA Goddard
Space Flight Center, Greenbelt, MD 20771, USA}
\altaffiltext{8}{CRESST, University of Maryland-College Park, College
Park, MD 20742, USA}
\altaffiltext{9}{NASA/Marshall Space Flight Center, ZP12, 320 Sparkman Drive, Huntsville, AL, 35805.}
\altaffiltext{10}{A.F.Ioffe Physical-Technical Institute, St. Petersburg 194021, also St.Petersburg State Politechnical University, Russia}

\begin{abstract}
We obtained six observations of PSR J1741-2054 using the $Chandra$
ACIS-S detector totaling $\sim$300 ks. By registering this new epoch
of observations to an archival observation taken 3.2 years earlier
using X-ray point sources in the field of view, we have measured
the pulsar proper motion at $\mu =109 \pm 10 {\rm mas\ yr}^{-1}$
in a direction consistent with the symmetry axis of the observed
H$\alpha$ nebula.  We investigated the inferred past trajectory of
the pulsar but find no compelling association with OB associations
in which the progenitor may have originated.  We confirm previous
measurements of the pulsar spectrum as an absorbed power law with
photon index $\Gamma$=2.68$\pm$0.04, plus a blackbody with an
emission radius of (4.5$^{+3.2}_{-2.5})d_{0.38}$ km, for a DM-estimated
distance of $0.38d_{0.38}$ kpc and a temperature of $61.7\pm3.0$
eV. Emission from the compact nebula is well described by an absorbed
power law model with a photon index of $\Gamma$ = 1.67$\pm$0.06,
while the diffuse emission seen as a trail extending northeast of
the pulsar shows no evidence of synchrotron cooling. We also applied
image deconvolution techniques to search for small-scale structures
in the immediate vicinity of the pulsar, but found no conclusive
evidence for such structures. \\
\end{abstract}

\keywords{pulsars: individual (PSR J1741-2054) - X-rays: general}

\section{INTRODUCTION}

PSR J1741$-$2054 (J1741) is one of the closest middle-aged ($\tau_{c}$
= 390 kyr) pulsars known. It has a period of $P$ = 413 ms and was
first discovered in $\gamma$-rays using the Large Area Telescope
(LAT) on the \textit{Fermi Gamma-ray Space Telescope} by a blind
search for periodic $\gamma$-ray pulsations from $Fermi$-LAT point
sources \citep{2009Sci...325..840A}.  It was subsequently detected
in archival Parkes radio data and observed using the Green Bank
Telescope \citep{2009ApJ...705....1C}. The pulsar has a spin down
energy loss rate of $\dot{E}$ = 9.5 $\times$ 10$^{33}$ erg s$^{-1}$
which is moderately low compared to those of other $\gamma$-ray
pulsars. The pulsar has a very small dispersion measure (DM) = 4.7
pc cm$^{-3}$ and a magnetic field of 2.7$\times10^{12}$G. Using the
NE2001 Galactic electron density model, the low DM implies a distance
of 0.38 kpc \citep{2002astro.ph..7156C}.  At this distance, its
measured radio flux at 1400 MHz ($S\sim160\mu Jy$) makes it one of
the least luminous radio pulsars known. Its $\gamma$-ray pulsations
lag behind its radio pulsations by $\delta$=0.29$P$, implying that
our line-of-sight tangentially cuts the $\gamma$-ray cone, while
nearly missing the radio beam \citep{2009ApJ...705....1C}. This
makes J1741 a transitional object between a classical radio/$\gamma$-ray
loud pulsar such as Vela and the radio-quiet Geminga-type pulsars.
Interestingly, \citet{2010ApJ...724..908R} detected a $20''$ long,
non-radiative H$\alpha$ bow shock nebula around the pulsar.  Modeling
of the bow shock suggested that the pulsar is traveling with a
velocity of $\sim$150 km s$^{-1}$, while the observation of negative
radial velocities from both sides of the nebula imply that the
velocity is directed out of the plane of the sky at an angle of
$15^{\circ}\pm10^{\circ}$.

\begin{figure*}[htbp!]
\begin{center}
\includegraphics[width=0.62\textwidth]{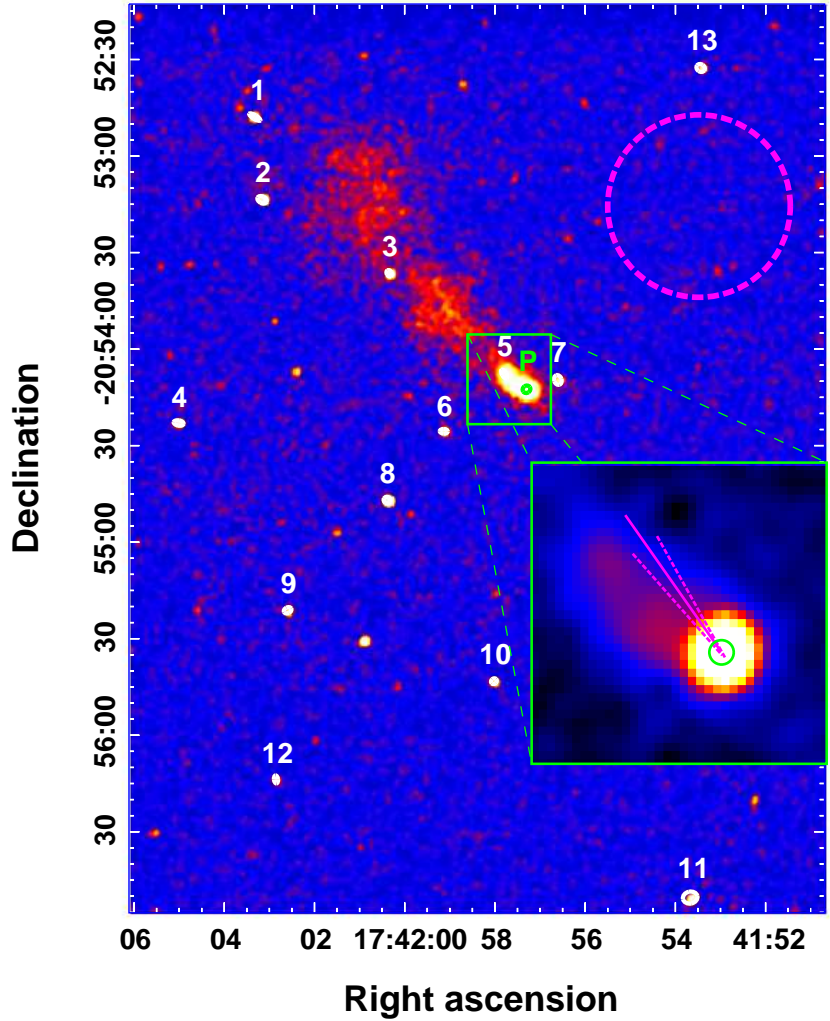}
\caption{
Merged $Chandra$ exposure-corrected 0.3-5.0 keV ACIS-S image of the
extended emission around J1741-2051. This was produced using
$reproject\_obs$ and $flux\_obs$ and incorporates all available
$Chandra$ observations. The image is smoothed using a Gaussian of
width $3''$ and plotted on the logarithmic scale. The reference
sources numbered 1 through 13 were used for relative astrometry,
while the pulsar is labeled as P. Other sources seen in the image
are not suitable for astrometry because, being variable, they were
not significantly detected in both the archival observation and one
of the new epoch observations. We included a cutout region around
the pulsar to show the point source. The magenta lines show our
derived proper motion (solid) and its uncertainty (dashed), traced
back from the pulsar position.  The background region used for
spectral analysis is shown as the magenta dashed circle. \label{PSRimage}
}
\end{center}
\end{figure*}

Using a short $Chandra$ ACIS-S observation
(observation ID (ObsID): 11251),  \citet{2010ApJ...724..908R}
detected an X-ray pulsar wind nebula (PWN) within this H$\alpha$
nebula and a long ($\textgreater$ 2 arcmin) X-ray trail at an angle
of $45^{\circ}\pm5^{\circ}$ East from North. \citet{2010ApJ...724..908R}
also suggested that there are asymmetries in the small scale structure
surrounding the pulsar, which they associate with a compact $2.5''$
equatorial toroidal structure. \citet{2014ApJ...790...51M} and
\citet{2014ApJ...789...97K} performed a spectral analysis of the
pulsar emission using \textit{XMM-Newton} and \textit{Chandra} and
determined that a two-component (blackbody plus power-law) model
is required to obtain satisfactory spectral fits.

In this paper, we use $\sim$300 ks of $Chandra$ data of J1741 that
were obtained as part of the Cycle 14 $Chandra$ Visionary Project
``A Legacy Study of the Relativistic Shocks of PWNe'', plus a
$\sim49$ks archival observation, to constrain the pulsar motion,
and the geometry of the PWN outflow. We discuss our approach to
image registration and proper motion measurement in Section 3,
followed by a discussion in Section 4 of our image deconvolution
efforts to search for small-scale structure around the pulsar. In
Section 5 we discuss the results of our spectral fits for the pulsar
emission and that of the extended PWN, and compare these with
previous results.  In Section 6 we discuss the implications of the
proper motion measurements, including comparisons with the observed
H$\alpha$ nebula surrounding J1741, and discuss the lack of
evidence of for synchrotron cooling in the PWN trail. Our conclusions
are summarized in Section 7.

\begin{figure}[htbp!]
\begin{center}
\includegraphics[width=0.45\textwidth]{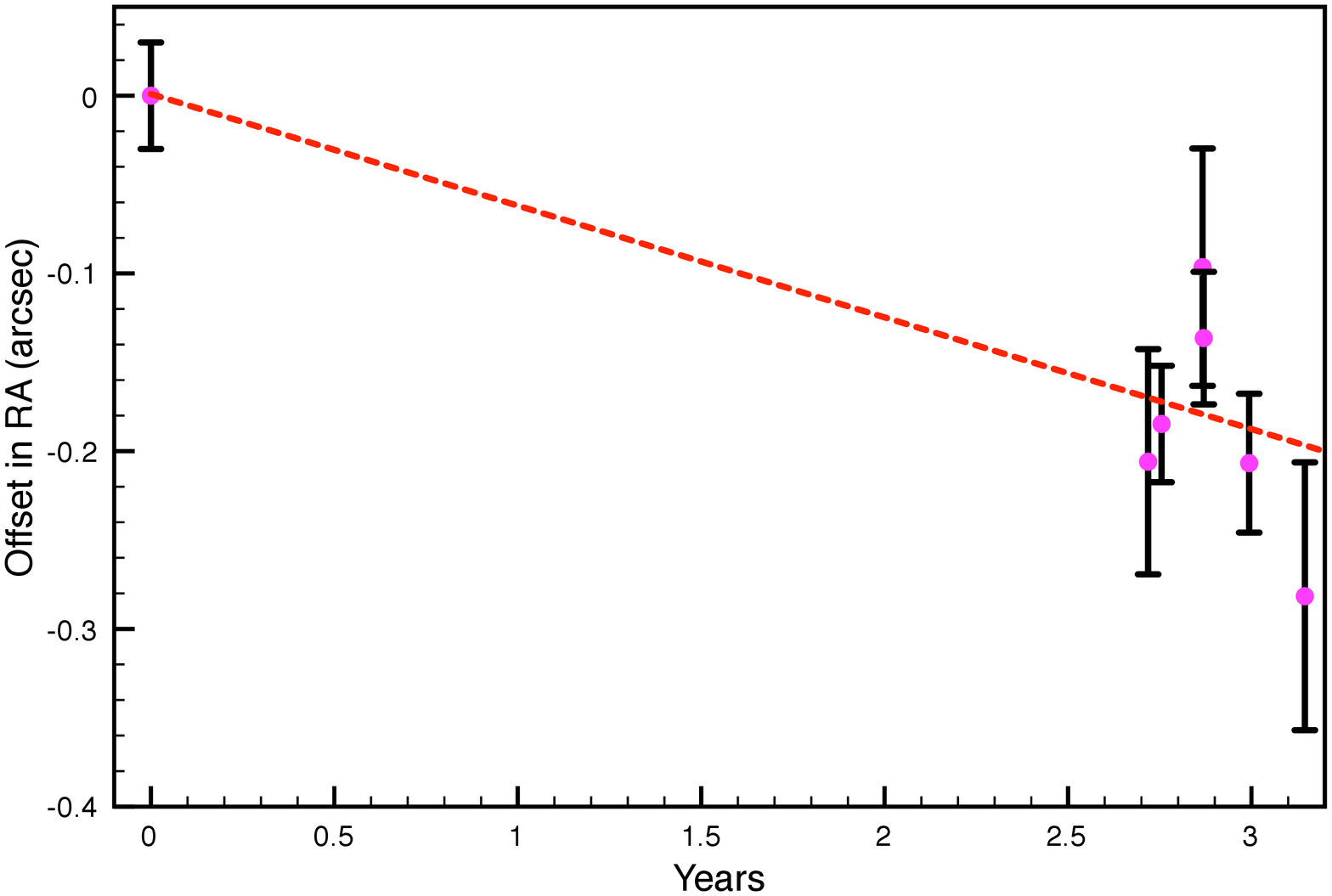}
\includegraphics[width=0.45\textwidth]{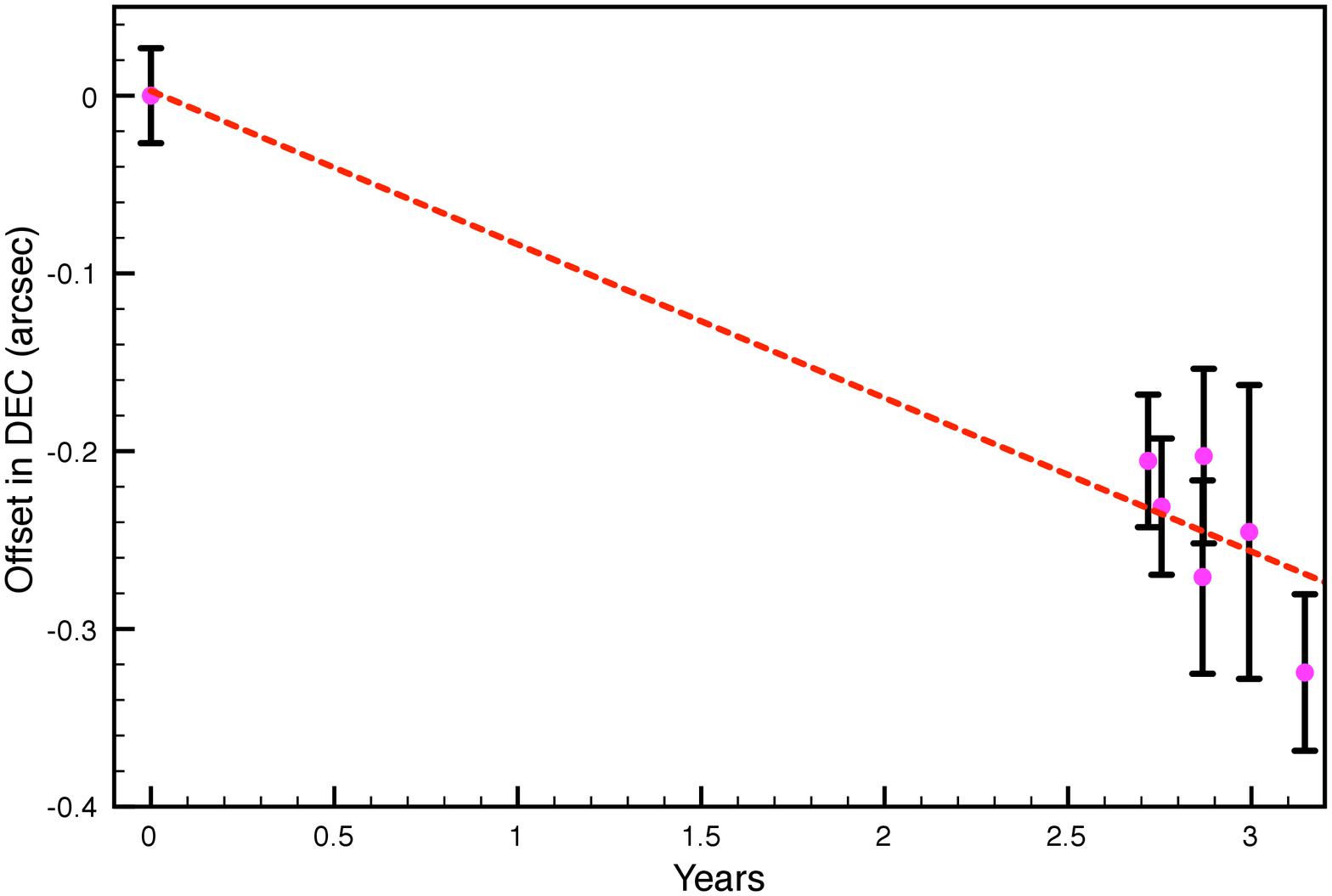}
\caption{
Offset in the position of the pulsar in RA (\textit{top}) and DEC
(\textit{bottom}) in the new observations from the position of the
pulsar in the archival observation plotted against the time since
the first observation of PSRJ1741 with $Chandra$ (years). The dashed
line corresponds to the line of best fit in which the slope corresponds
to the proper motion of the pulsar.\label{offset}
}
\end{center}
\end{figure}

\section{X-ray Data Analysis}

We obtained 282 ks of new $Chandra$ ACIS-S exposure time of J1741. Table
\ref{obsids} lists the parameters for each of the six new observations
that we obtained, including the archival observation. To reduce
pileup, the CCDs were operated in half-frame VFAINT mode so that
events were read out every 1.7s. The pulsar was placed near the
optimum focus on the backside illuminated S3 chip. In addition to
our six observations, there is a 48.8 ks archival observation (ObsID:
11251), which was taken on 2010 May 21. Each new observation has a
roll angle similar to the roll angle of the 2010 epoch ($\sim$
90$^{\circ}$), except for ObsID: 15544. This observation has a roll
angle of $\sim$ 260$^{\circ}$ as the nominal roll angle of $Chandra$
is rotated by 180$^{\circ}$ during the time of the year this
observation was completed. The data were analysed with $CIAO$ 4.6.2
after all observations were reprocessed using the CALDB 4.5.9. No
flaring occurred in any of the observations so the full exposure
times were used.

Using all seven observations we produced a merged, exposure-corrected
image of J1741 by reprojecting the new observations to a common
tangent plane based on the WCS information of ObsID: 11251 ($CIAO$
task: $reproject\_obs$) and combined all reprojected observations
into an exposure corrected image using the $CIAO$ task $flux\_obs$.
The merged ACIS-S image of the extended emission around the pulsar,
smoothed with a $3''$ Gaussian", is shown in Figure \ref{PSRimage}.
The pulsar point source lies at the apex of the diffuse X-ray
emission, while a diffuse, faint X-ray trail extending $\sim$ $1.9'$
is seen towards the northeast of the pulsar.

\begin{table}[t!]
\begin{center}
\caption{$Chandra$ observations of PSR J1741-2054 \label{obsids}}
 \begin{tabular}{ccc}
Observation ID & Exposure Time (ks) & Observation Date \\
\hline\hline
$archival$ &  &  \\
11251 & 48.78& 2010-05-21\\
\textit{new observations}&  &  \\
14695 & 57.15 & 2013-02-06 \\
14696 &54.30& 2013-02-19 \\
15542 & 28.29 & 2013-04-01 \\
15638 & 29.36 & 2013-04-02 \\
15543 &57.22 & 2013-05-15 \\
15544 & 55.73 & 2013-07-09\\
\hline
\end{tabular}
\end{center}
\end{table}

\section{REGISTRATION AND THE PROPER MOTION}
To constrain the proper motion of the pulsar, we registered each
of the new Chandra images to the archival image using field point
sources that were identified using the $CIAO$ tool $wavdetect$. We
selected sources with a detection significance of $\textgreater
3\sigma$ that were found on the S3 chip in both the 2010 observation
and the corresponding new observation. These sources are highlighted
in white and the pulsar is labeled as P in Figure \ref{PSRimage}.

Careful consideration and modeling of the point-spread function
(PSF) must be undertaken to reduce the effect of changes in the PSF
shape on the count distribution of our point-sources. To improve
the astrometric accuracy and reduce this effect, we simulate a PSF
for each point source position, in each observation, and use it to
fit for the position of the source.

To simulate a PSF of a point source, we use the software suite
$SAOTrace$\footnote{\url{http://cxcoptics.cfa.harvard.edu/SAOTrace/Index.html}}
which is designed to simulate the propagation of photons from
astronomical objects through the optics of the $Chandra$ X-ray
satellite. We use the aspect solution file of each observation and
provided spectral information for the ray trace by extracting the
spectrum of each stellar source using the $CIAO$ task $specextract$.
We model the spectra using an absorbed Mekal model in $SHERPA$. To
improve the accuracy of the PSF modeling, we increase the normalisation
of this spectrum by a factor of 100 before passing it into $SAOTrace$.
A model of a point source at its position is obtained by passing
the raytrace from $SAOTrace$ into the program
$MARX$\footnote{\url{http://space.mit.edu/CXC/MARX/}}. 
Each PSF model is corrected for the science instrument
module (SIM) offset from nominal location and filtered using the
Good Time Interval (GTI) data from the original event file.

To determine the position of the sources we use the maximum likelihood
``figure of merit'' (FoM) technique developed by
\citet{2012ApJ...755..151V}.  We generate a 39 pixel by 39 pixel
image of the modeled PSF, binned to 1/9 ACIS pixel resolution. This
is then compared to 0.3 - 5.0 keV source images of the same size
but binned to native ACIS pixel resolution. PSF models and source
images are produced for all observations and for each registration
source. The PSF is shifted along the $x$ and $y$ axes of the 1/9
pixel grid and rebinned to native ACIS pixels. We then compute the
FoM at each offset in pixel coordinates, giving us a map of the
likelihood of the observed counts with respect to the $x$ and $y$
position. To determine the best-fit position of the source, we fit
a two dimensional Gaussian to the FoM surface, with the minimum of
this surface providing the best fit position of the source. The
standard deviations along $x$ and $y$ are estimated by calculating
the square root of the eigenvalues of the covariance matrix.

\begin{figure}[tbp!]
\begin{center}
\includegraphics[width=0.5\textwidth]{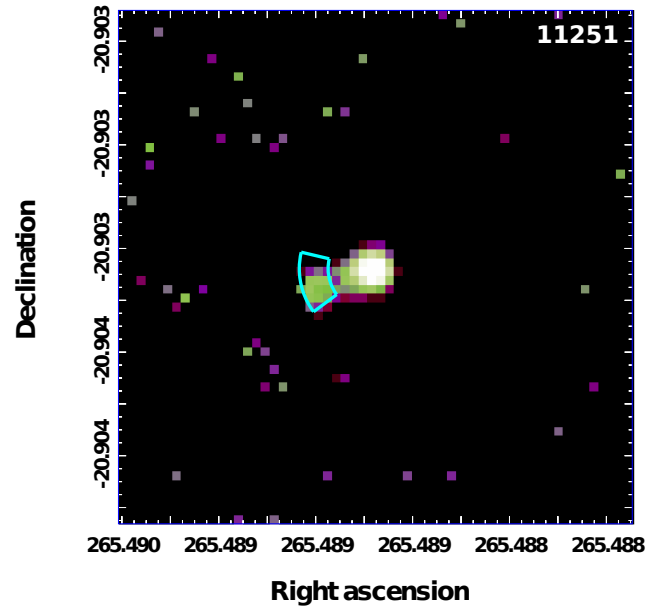}
\caption{
Deconvolved image of J1741-2054 from ObsID 11251 obtained using
$arestore$ after 50 iterations. The cyan region corresponds to the
PSF asymmetry seen in $Chandra$ data when pushing to sub-ACIS-pixel
resolution. The image has been logarithmically stretched and the
colour enhanced to highlight the observed features.
\label{arestore}
}
\end{center}
\end{figure}

\begin{figure*}[t!]
\begin{center}
\includegraphics[height=0.30\textheight]{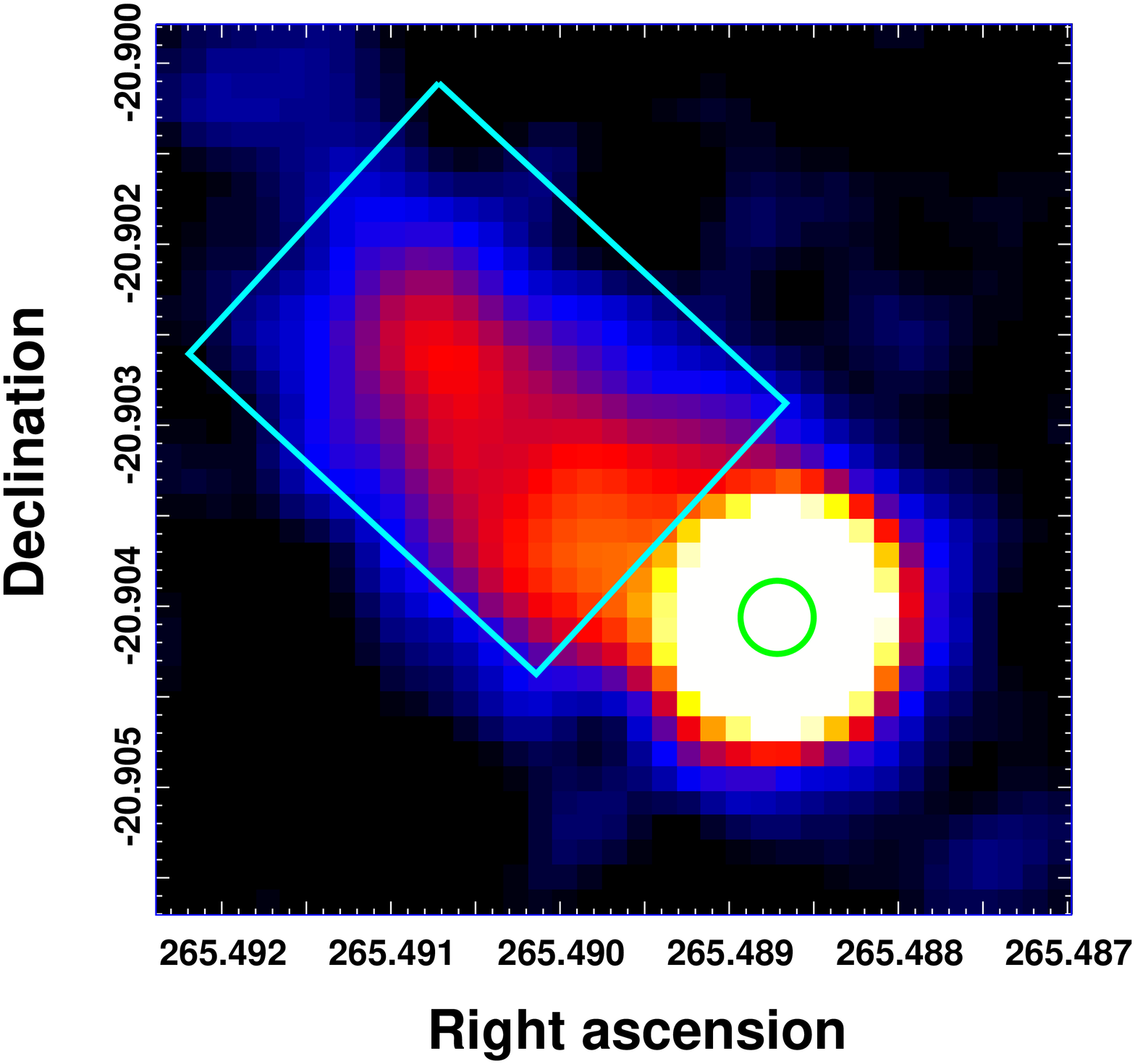}
\includegraphics[height=0.30\textheight]{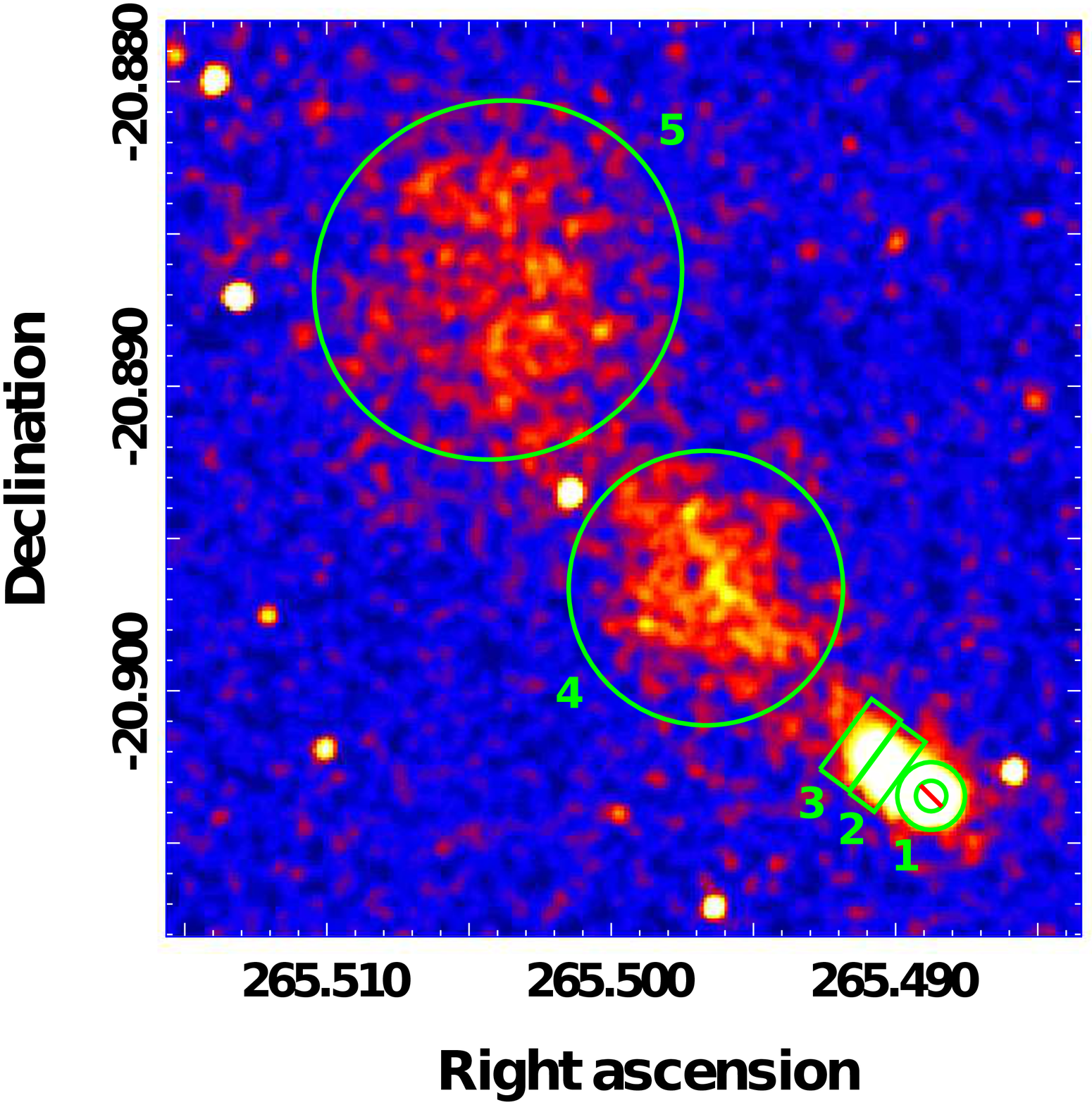}
\caption{
Logarithmically scaled, merged $Chandra$ exposure-corrected 0.3-5.0
keV ACIS-S image of the pulsar and its extended emission. $Left$:
The green circular region defines the $0.70''$ region used to extract
the spectrum of the pulsar from all available observations. The
cyan rectangular region defines the region used to extract the
spectrum of the pulsar's compact nebula from all observations.
$Right$: The green regions define the apertures used to look for spectral
variability in the extended emission of the pulsar.  The pulsar was
excluded from the spectral extraction of region 1.\label{specextract}
}
\end{center}
\end{figure*}

Prior to registration we checked each source for any optical
counterparts using
VizieR\footnote{\url{http://vizier.u-strasbg.fr/viz-bin/VizieR}}.
Sources 6, 8, 9, 10 and 13 all have optical counterparts within
$2''$ of the PSF-fit position.  Source 8 and 13 have estimated
proper motions of ($\mu_{RA},\mu_{DEC}$) = (22,$-$62) ${\rm mas\
yr}^{-1}$ and (20,-22) ${\rm mas\ yr}^{-1}$, respectively (errors
$\delta\mu_{RA} \sim 7$ ${\rm mas\ yr}^{-1}$, $\delta\mu_{DEC} \sim
$15 ${\rm mas\ yr}^{-1}$). We corrected for these nominal proper
motions, but also confirmed that our final astrometric solution was
insignificantly changed if we excluded these two stars from the
analysis.

Using the FoM positions of our registration sources and their
uncertainties as a reference grid to perform the relative astrometry,
we determined the best translation transformation needed to register
the images from the two epochs using the $CIAO$ command
$wcs\_match$.  For each observation we adopt the $Chandra$-determined
roll angle. In Table \ref{shifts} we list the best-fit frame shifts
and their uncertainties. The uncertainties in these shifts were
calculated by adding in quadrature the errors in the differences
between each source before and after shift. Adding a rotation to
the transformation did not produce a statistically significant
improvement on the best-fit translation.

\begin{table}[t!]
\begin{center}
   \caption {Frame shifts and their uncertainties used for
registration.\label{shifts}}
   \begin{tabular}{ccccc}
ObsID   &       RA              &       DEC                     \\
        &       (arcsec)                        &       (arcsec)        \\
        \hline \hline
14695   &       $-$0.02 $\pm$   0.02    &       0.48    $\pm$   0.02    \\
14696   &       $-$0.01 $\pm$   0.02    &       0.25    $\pm$   0.02    \\
15542   &       0.02    $\pm$   0.04    &       0.43    $\pm$   0.04    \\
15638   &       $-$0.06 $\pm$   0.03    &       0.36    $\pm$   0.03    \\
15543   &       0.23    $\pm$   0.02    &       0.29    $\pm$   0.03    \\
15544   &       $-$0.11 $\pm$   0.03    &       $-$0.29 $\pm$   0.03    \\
\hline
\end{tabular}
\end{center}
\end{table}

To calculate the position of the pulsar after registration, we first
calculate the position of the pulsar in the unregistered frames.
We simulate a PSF at the position of the pulsar in each observation
using the method described  for the registration sources. To define
the energy dependence of the PSF, we extract a spectrum of the
pulsar and fit it with an absorbed power law plus blackbody model
(see Section 5). For each observation, we generate a 6 pixel by 6
pixel image of the PSF model of the pulsar that is binned to 1/4
ACIS pixel resolution. Using an image of the 0.3 - 5.0 keV pulsar
events of the same size and binning, we fit for the position of the
pulsar in $SHERPA$. In this fit, we used a delta function (the
pulsar) plus a two dimensional Gaussian (the circumpulsar PWN),
both convolved with the PSF. The resulting pulsar fit position for
each frame was then registered by applying the best-fit transformations
(Table \ref{shifts}).

\begin{table*}[t!]
\begin{center}
   \caption {Spectral fit values of the extended emission of the pulsar
defined by the cyan coloured region in Figure \ref{specextract}a and the
green regions defined in Figure \ref{specextract}b. All uncertainties are one
$\sigma$.\label{trail}}
    \begin{tabular}{cccccc}
    \hline
    Region   & $N_{H}(\times 10^{21}$)cm$^{-2\ast}$  & $\Gamma$
& Absorbed Flux   & Unabsorbed Flux                   & $\chi^{2}/dof$ \\
    ~    & ~           & ~    & $10^{-14}$ erg cm$^{-2}$ s$^{-1}$ &
$10^{-14}$ erg cm$^{-2}$ s$^{-1}$ & ~               \\
\hline  \hline
compact PWN &$1.20^{+0.08}_{-0.07}$ & $1.60\pm0.20$
&$2.86^{+0.17}_{-0.20}$&$3.15^{+0.09}_{-0.07}$&0.90\\
     1    & - & 1.97$^{+0.18}_{-0.17}$          & 1.56$^{+0.13}_{-0.12}$
& 1.81$\pm0.33$        & 0.99         \\
    2    &  -& 1.50$^{+0.16}_{-0.15}$& 1.54$^{+0.15}_{-0.10}$       &
1.67$\pm0.16$        & 0.84 \\
    3    &  - & 1.57$^{+0.20}_{-0.19}$          & 0.81$^{+0.08}_{-0.11}$
& 1.10$\pm0.11$        & 0.80        \\
    4    & -& 1.63$^{+0.12}_{-0.11}$ & 4.35$^{+0.28}_{-0.15}$        &
4.81$^{+0.04}_{-0.14}$        & 0.99         \\
    5    &   -      & 1.70$^{+0.10}_{-0.11}$  & 5.40$^{+0.29}_{-0.36}$   &
6.03$^{+0.06}_{-0.16}$   & 1.16          \\
 \hline
$^{\ast}$ \footnotesize{Fixed at value from joint compact nebula.}\\
    \end{tabular}
 \end{center}

 \vspace{-3mm}

\end{table*}

To quantify the proper motion of the pulsar, we plot in Figure
\ref{offset} the offset in the position of the pulsar between the
archival observation and the new observations against the number
of years since the archival observation. The uncertainties in the
offset are calculated by adding in quadrature the uncertainties in
the fit positions of the pulsar, the uncertainty in the frame shifts
and the systematic uncertainty associated with choosing a particular
tangent plane when creating an image in sky coordinates. We fit the
offset using a linear function that corresponds to the positional
shift of the pulsar between the archival observation and the new
observations and this is seen as the dashed line in Figure \ref{offset}.

We obtain a proper motion of $\mu_{RA}cos(\delta) = -63 \pm 12$
${\rm mas\ yr}^{-1}$ and $\mu_{DEC} = 89 \pm 9$ ${\rm mas\ yr}^{-1}$.
This corresponds to a total proper motion of 109$\pm$10 ${\rm mas\
yr}^{-1}$. Assuming a distance of 0.38 kpc to the pulsar, this
translates to a transverse velocity of $(196\pm18)d_{0.38}$ km
s$^{-1}$. The position angle of the proper motion is
$215^{\circ}\pm6^{\circ}$ east of north. The proper motion axis
points in the opposite direction of the extended X-ray trail as
expected (see Figure \ref{PSRimage}).

\section{X-RAY IMAGING}

Using the archival $Chandra$ observation of J1741 (ObsID: 11251),
\citet{2010ApJ...724..908R} performed a PSF subtraction of the
pulsar point source to look for any small-scale structure surrounding
the pulsar. They discovered that the region around the pulsar appears
to be slightly extended and they associate this feature with the
equatorial torus of the PWN. Using the same data as analyzed here,
\citet{2014ApJ...789...97K} searched for evidence of such structure by
performing fits to a two-dimensional Gaussian convolved with PSF
models generated for each observation. They find no evidence for
any small-scale extended features other than for a small emission
feature associated with a known mirror
artifact\footnote{\url{http://cxc.harvard.edu/ciao/caveats/psf\_artifact.html}}.
We have carried out a similar investigation using image deconvolution
techniques.  We simulate a PSF of the pulsar in each observation
using MARX following re-reduction of our $Chandra$ observations
with CALDB 4.4.7 to match the calibration data used for MARX.  We
define the energy dependence of the PSF as described in Section 3
and use the dither pattern of the observation with an aspect blur
of 0.07, which corresponds to the uncertainty in the telescope
pointing\footnote{\url{http://space.mit.edu/CXC/marx/news.html}}.
We also correct for SIM offset. Using this PSF we deconvolve a 0.3
- 5.0 keV pulsar image using the Lucy-Richardson deconvolution
algorithm \citep{1974AJ.....79..745L} implemented using the $CIAO$
task $arestore$. All images were binned to quarter-ACIS pixel. We
ran $arestore$ multiple times using a number of different iterations
between 10 and 200 to determine convergence of new features. No new
structures appeared after 50 iterations.

The deconvolved image from one observation is shown in Figure
\ref{arestore}.  The emission is well described by a point source,
with the exception of the artifact feature (identified in cyan).
We thus see no conclusive evidence of other small-scale structure
in the immediate region surrounding the pulsar, consistent with the
results reported by \citet{2014ApJ...789...97K}.

\begin{figure*}[htbp!]
\begin{center}
\includegraphics[width=0.8\textwidth]{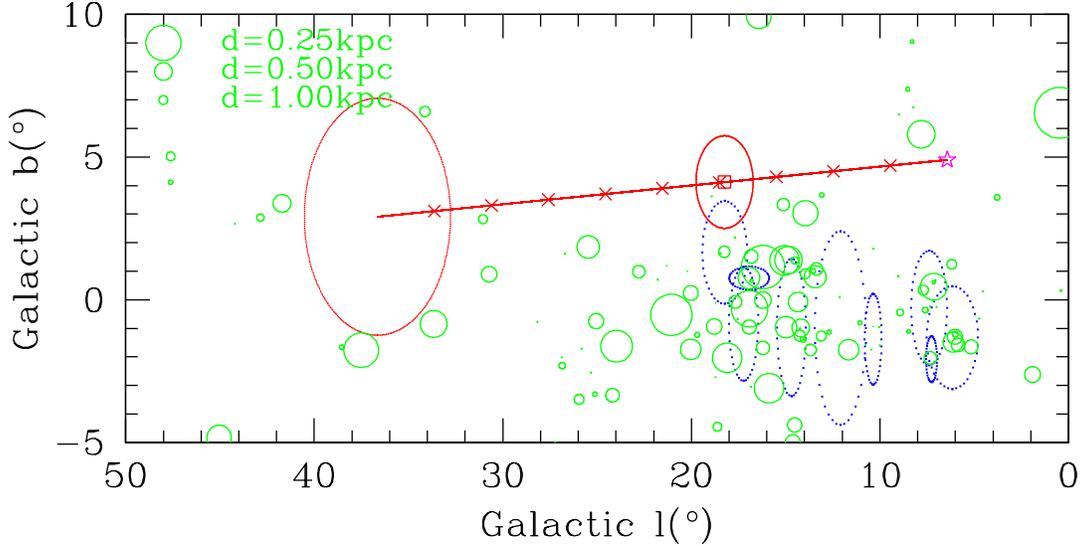}
\caption{
The past trajectory (red) of PSR J1741$-$2051 in Galactic coordinates.
Positions are marked every $10^5$y, with uncertainty ellipses at
$\tau_c = 3.91 \times 10^5$y and at $10^6$y. For comparison we show
the locations and sizes of OB associations from the \citet{me95}
catalog (blue dotted ellipses) and the Hipparcos sample of OB stars
(green circles). A substantial concentration of OB stars at $\sim
0.3-0.6$\,kpc lies near the pulsar track at $\sim \tau_c$ and a
more distant ($d=1.45$\,kpc) OB association overlaps with the track
uncertainty at this age.
}
\label{OB}
\end{center}
\end{figure*}

\section{SPECTRAL ANALYSIS OF THE X-RAY EMISSION OF THE PULSAR AND ITS TRAIL}
\subsection{Pulsar}

As noted above, modeling of the $Chandra$ PSF for use in determining
an accurate pulsar position requires knowledge of the source spectrum.
\textit{XMM-Newton} observations establish a two-component spectrum
for the pulsar, with a blackbody accompanied by a power law
\citep{2014ApJ...790...51M}.  Similar results were derived by
\citet{2014ApJ...789...97K} using the same $Chandra$ data reported
here. We have re-analyzed these data by extracting events from a
$0.70''$ radius region centered on the pulsar, shown as the green
circle in Figure \ref{specextract}a.  The size of this region was
chosen to minimise the contamination from the PWN, and subsequent
modeling results have been corrected for the finite encircled energy
fraction.  All spectra were grouped with a minimum of 20 counts per
bin, and a background spectrum was obtained using a source-free
circular region with a radius of $30''$, as shown by the magenta
circle in Figure \ref{PSRimage}.  Using the $CIAO$ $pileup\_map$,
we determined that approximately 5\% of the pulsar events suffer
from pileup. We thus included a pileup model in our spectral fits,
where the frame time is 1.7s and the PSF fraction is allowed to
vary. All other parameters in the pileup model were frozen at default
values.

To constrain the column density, we fit spectra from the compact
nebula to the northeast of the pulsar (see Section 5.2)  using an
absorbed power law (more details in section 5.2). We obtained $N_{H}$
= $(1.20^{+0.08}_{-0.07})\times10^{21}$cm$^{-2}$, in good agreement
with the above studies, and adopt this value in all our models of
the pulsar spectrum. We use the Wilms et al. abundance table throughout
our analysis \citep{2000ApJ...542..914W}.  Modeling the spectrum
with a power-law plus blackbody model, we obtained a photon index
of $\Gamma = 2.68\pm0.04$ and a blackbody spectrum with a temperature
of $kT_{\rm eff} = 61.7\pm3.0$ eV, in excellent agreement with the
results reported by \cite{2014ApJ...790...51M} and
\citet{2014ApJ...789...97K}.  Omitting the pile-up correction yields
similar values. We use the best-fit values above for PSF modeling
of the pulsar.

We also ran fits using magnetised neutron star atmosphere models
(\textit{nsa} and \textit{nsmax} in SHERPA) for the thermal component.
These gave somewhat different temperatures and emitting areas, but
did not significantly improve the quality of the fit. For example
a magnetic carbon atmosphere model (\textit{nsmax} model 12006,
\citet{2007MNRAS.377..905M}) gave a temperature of $kT_{eff}=86.0\pm9.0$
eV and emission radius of $R_{emis}=(4.90^{+3.0}_{-2.3})d_{0.38}$
km. The power law component was only slightly affected with
$\Gamma=2.63\pm0.03$.

\subsection{PWN and its extended X-ray trail}
To analyse the spectrum of the compact nebula described above, we
extracted spectra from each observation using the cyan rectangular
region in Figure \ref{specextract}a and combined these using
$specextract$. We used the same background spectrum as for the
pulsar spectrum. The compact nebula contains a total of $\sim$900
counts and we binned the combined spectrum with a minimum of 20
counts per spectral bin. The PWN spectrum is consistent with an
absorbed power law with an index of $\Gamma =1.60 \pm 0.20$. 

To determine whether there is any spectral variation in the PWN and
its extended trail, we extract spectra from the five regions defined
in Figure \ref{specextract}b. These regions correspond to roughly
the same spectral regions reported by \citet{2014ApJ...790...51M},
except that we investigate smaller regions in the compact nebula
near the pulsar.  In the outer portions of the nebula, the count
rate is too low to obtain good spectra in smaller regions. We model
each region individually using an absorbed power law, where we fix
column density to the value derived earlier from fits to the inner
nebula region but let photon index and the normalisation vary. We
have listed in Table \ref{trail} the absorbed flux, unabsorbed flux
and the reduced $\chi^{2}$ for each region, as well as the best-fit
parameters from fitting the trail spectra.  There is no evidence
of systematic variation in the photon index of the compact nebula
and trail (regions 1 - 5). The photon index of region 1 is slightly
higher than that of the other four regions, but is consistent within
uncertainties with all regions except for region 2. This slight
variation between region 1 and 2 could suggest that region 1 is
affected by leakage of the softer emission from the point source.
Modeling the photon index as a function of distance from the pulsar
using a linear regression fit with a constant function in $SHERPA$,
we obtain $\Gamma = 1.67 \pm 0.06$ for the trail. The global PWN
index derived from the $SHERPA$ fit and the values we obtained in
Table \ref{trail} are consistent with \citet{2014ApJ...789...97K}
and \citet{2014ApJ...790...51M}, who derived $\Gamma=1.74\pm0.07$
and $\Gamma=1.78\pm0.15$ respectively for the PWN.

\section{DISCUSSION}

Using $Chandra$ observations of PSR J1741-2054 that span $\sim$ 3.2
year period, relative astrometry measurements have identified a
proper motion of $\mu$=109$\pm$10 ${\rm mas\ yr}^{-1}$. This
corresponds to a modest velocity of $(196\pm18)d_{0.38}$ km s$^{-1}$,
which agrees well with the velocity derived by \citet{2010ApJ...724..908R}
using H$\alpha$ spectroscopy. The larger distance of
\citet{2014ApJ...789...97K} gives a transverse velocity of $\sim
400$ km/s, inconsistent with that obtained from optical spectroscopy.
The direction of the proper motion is 205$^{\circ}\pm$6$^{\circ}$
east of north, opposite the elongated X-ray trail.
\citet{2004ApJ...601..479N}, \citet{2005MNRAS.364.1397J} and
\cite{2007ApJ...660.1357N} found that the direction of proper motion
of a pulsar is, generally, approximately parallel to its rotation
axis.

In Figure \ref{OB}  we plot the pulsar track (red) in Galactic
coordinates (note the expanded $b$ scale).  This has the pulsar
skimming above the plane. It does not intersect the plane itself
unless one extrapolates an unreasonable $\sim 2 \times 10^6$y;
however, the track starts within the $\sim 50$\,pc OB star scale
height for $t \sim \tau_c$ and distances $d<1$\,kpc. For comparison
we plot the positions of the Hipparcos catalog OB stars (green),
with circle size proportional to the parallax. This set is quite
complete, with useful parallaxes, to $\sim 500$\,pc, and increasingly
incomplete at larger distances.  At large distance the cataloged
OB associations \citep{me95} provide plausible pulsar birthsites,
and their cataloged extent is plotted by the blue dotted ellipses.
Intriguingly, one association overlaps the pulsar track, but this
is at a likely unreasonable catalog distance of 1.45~kpc. We conclude
that, with the pulsar motion passing along the Galactic plane, there
will be many superposed massive star locations, and no definitive
birthsite can be identified.  However, there are certainly many
plausibly associated massive stars consistent with our preferred
$d\sim 0.4$\,kpc, especially considering that some pulsar progenitors
may be OB runaways with significant offset during their pre-explosion
lifetime.

\begin{figure}[tb!]
\begin{center}
\includegraphics[width=0.40\textwidth]{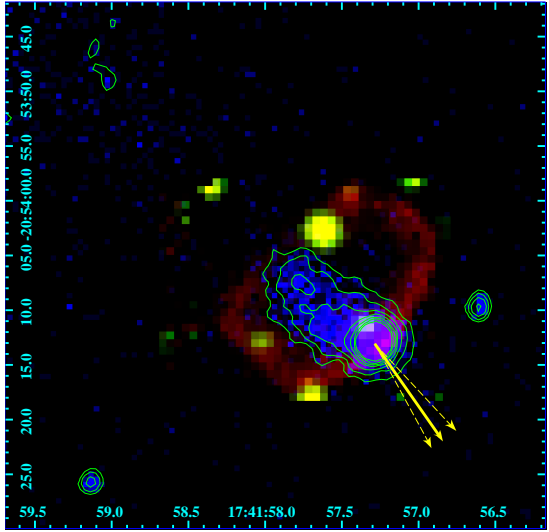}
\caption{
The head of the PSR J1741$-$2051 nebula. The red channel shows H$\alpha$
emission at $6562-6564$\AA\ while the green channel shows average
of the continuum $6551-6561, 6565-6575$\AA\, with data abstracted
from an AAT SPIRAL IFU observation. The blue channel show the
$0.3-7$keV X-ray photons from our combined new observation.  Green
contours help one to read the X-ray structure, while the yellow
arrows show our measured proper motion, with the arrow heads marking
the position in 100 yr.
}
\label{halpha}
\end{center}
\end{figure}

Neither our deconvolved nor our PSF-subtracted images indicate
conclusive evidence of small-scale structure surrounding the pulsar
that might be associated with a torus or jet-like feature. The
equatorial torus structure that \citet{2010ApJ...724..908R} associate
with a diagonal excess seen $\sim 0.75''$ from the core of the
pulsar image seems to have arisen from the mirror asymmetry.
\citet{2014ApJ...789...97K} perform a similar analysis and come to
the same conclusion.  However, it is interesting to compare the
nebula head and proper motion with the H$\alpha$ structure described
in Romani et al (2010).  In Figure \ref{halpha} we see that the
pulsar lies very close to the bow shock limb (accuracy limited by
our relative X-ray/optical astrometry).  Interestingly, our measured
proper motion is consistent with, although nominally slightly south
of, the H$\alpha$ nebula's symmetry axis. However the X-ray PWN
trail fills only the southern half of the apparent H$\alpha$ cavity,
punching out through a gap at the back end of the H$\alpha$ emission
and continuing to the arcmin-scale trail beyond. The origin of this
asymmetry is unclear, but a clue may be seen in the X-ray contours,
whose ridge line lies at PA$\approx 70^\circ$, i.e., misaligned
with the proper motion by $\approx 35^\circ$. This suggests a second
symmetry axis in the PWN, possibly due to a pulsar jet or other
outflow concentration.  This directs the shocked PWN plasma to the
southeast, preferentially filling this half of the H$\alpha$ cavity.
A more complete discussion of the PWN geometry, including the 3D
H$\alpha$ kinematics, is in preparation.

The X-ray spectrum of the pulsar requires a
combination of non-thermal and thermal model components. The emission
is dominated by the non-thermal component ($\sim$ 75\% of unabsorbed
flux), indicating that the majority of the X-ray emission is
magnetospheric in nature.  The emitting radius implied by the
blackbody model for J1741 corresponds to (4.5$^{+3.2}_{-2.5})d_{0.38}$
km. This is substantially smaller than any viable neutron star
radius \citep{2007PhR...442..109L}, suggesting that this thermal
emission arises from hot spots on the surface, plausibly near the
magnetic poles \citep{2009Natur.462...71H}. In fact,
\citet{2014ApJ...790...51M} do detect a pulsed thermal component
for PSR J1741, also supporting such a surface temperature inhomogeneity.

The X-ray emission from the compact nebula and the trail is consistent
with an absorbed power law. There is no discernible evidence of
spectral variation with distance from the pulsar and the spectrum
of the entire tail can be described by $\Gamma = 1.67 \pm 0.06$.
We compute the minimum (equipartition) energy by approximating the
X-ray emission from the PWN (region 2, 3, 4 and 5 in Figure
\ref{specextract}b) as a cylinder with length $l \sim$108 arcsecs
($0.20 d_{0.38}{\rm\ pc}$) and width $w \sim$18 arcsecs ($0.03
d_{0.38}{\rm\ pc}$), comprising a volume of $V \sim 5.0\times
10^{51} \phi d_{0.38}^{3}$ cm$^{3}$, where $\phi$ is filling factor.
The minimum energy in relativistic particles and magnetic field
required to produce a synchrotron source of a given luminosity
\citep{1970ranp.book.....P} yield $E_{min} \sim C (1+\kappa)^{4/7}
V^{3/7} L_{syn}^{4/7}$, where $\kappa$ is the ion to electron energy
ratio, $L_{syn}$ is the synchrotron luminosity and $C$ is a function
dependent on energy, electron charge, speed of light and the mass
of the electron in Gaussian cgs units (see \citet{1970ranp.book.....P}).
In the following, we considering only the leptonic case, where
$\kappa=0$. The total luminosity of the PWN is $L$(0.5-10.0 keV) =
$2.36\times10^{30}$ erg s$^{-1}$, giving $E_{min} \sim 5.50\times
10^{40} \phi^{3/7} d_{0.38}^{17/7}$ erg. The associated minimum-energy
magnetic field is $B_{min} \sim (D (1+\kappa) L_{syn})^{2/7}
V^{-2/7}$, where D is a function similar to $C$. This magnetic field
is $\sim 15 \phi^{-2/7} d_{0.38}^{-2/7}\mu$G, leading to a lifetime
of the X-ray emitting leptons of $\tau_{syn} \sim 6.4 \times 10^{4}
B_{\mu G}^{3/2} E_{keV}^{-1/2}$ years or $\sim 1100$yr at an observed
photon energy of 1 keV. This is comparable to the (length/proper
motion) = $108''/109 \sim 10^{3}$ yr required for the pulsar to
traverse the bright trail with our observed proper motion. Thus it
is not surprising that there is no dramatic spectral steeping along
the trail. If the PWN electrons flow at even faster speeds within
the trail, this conclusion is even stronger.

\section{CONCLUSION}

Using $\sim 300$ ks of $Chandra$ ACIS-S observations of PSR J1741-2054,
we were able to determine the proper motion of the pulsar with a
detection significance $>3\sigma$. The direction of the proper
motion is aligned with the extended PWN emission, and corresponds
well with a symmetry axis of the associated H$\alpha$ nebula. The
diffuse X-ray emission immediately behind the pulsar is concentrated
in the southeastern portion of the H$\alpha$ nebula, possibly
suggesting another flow axis from a jet or torus in the pulsar
system. The trajectory of the pulsar, extrapolated over the
characteristic age, does not provide a compelling correlation with
known OB associations at the distance of the pulsar, although there
are many massive stars consistent with this distance that could
potentially have had a common origin.

The pulsar spectrum is well described by an absorbed power law
accompanied by a blackbody with an emission radius of
(4.5$^{+3.2}_{-2.5})d_{0.38}$ km and a temperature of $kT_{eff}=61.7\pm3.0$
eV, as found in earlier works. The thermal component, a hot region
on the neutron star surface, is augmented by a magnetospheric or
unresolved PWN power law component. The PWN plus its extended trail
can be well described using an absorbed power law and there is no
evidence of variation in the photon index with distance from the
pulsar. The integrated luminosity of the PWN over the 0.5 - 10 keV
is $2.36\times10^{30}$ erg s$^{-1}$. This represents 0.02\% of the
pulsar spin down power, which is not atypical. We find no conclusive
evidence of small-scale structure surrounding the pulsar that we
can associate with a torus or jet-like structure.

\bibliography{mybib}

\end{document}